\documentclass[12pt]{revtex4}

\usepackage{amsmath}
\usepackage{amsfonts}
\usepackage{amssymb}
\usepackage{graphicx}

\newcommand{\df}{\mathrm{d}}

\newcommand{\im}{\mathit{i}}
\DeclareMathOperator{\tr}{Tr}

\begin{document}

\title{Inhomogeneous states in two dimensional linear sigma model  at large $N$}

\author{A.\ Pikalov$^{1,2}$}

\email{arseniy.pikalov@phystech.edu}

\affiliation{ $^{1}$\mbox{Moscow Institute of Physics and Technology, Dolgoprudny 141700,
Russia} \\
 $^{2}$Institute for Theoretical and Experimental Physics, Moscow,
Russia}

\date{\today}

\begin{abstract}
In this note we consider inhomogeneous solutions of two-dimensional linear sigma model in the large $N$ limit. These solutions are similar to the ones found recently in two-dimensional $CP^N$ sigma model. The solution exists only for some range of coupling constant. We calculate energy of the solutions as function of parameters of the model and show that at some value of the coupling constant it changes sign signaling a possible phase transition. The case of the nonlinear model at finite temperature is also discussed. The free energy of the inhomogeneous solution is shown to change sign at some critical temperature.
\end{abstract}

\maketitle

\section{Introduction}
Two-dimensional linear sigma model is a theory of $N$ real scalar fields and quartic $O(N)$ symmetric interaction.
The model has two dimensionful parameters: mass of the particles and coupling constant. 
In the limit of infinite coupling one can obtain the nonlinear $O(N)$ sigma model.
For that reason the linear model can be thought as a generalization of the nonlinear one.
These models can be solved in the large $N$ limit, see 
\cite{Novikov1984} for a review.
In turn $O(N)$ sigma model is quite similar to the  $CP^N$ sigma model. 

Recently the large $N$ $CP^N$ sigma model was considered on a finite interval with various boundary conditions \cite{Milekhin2012, Milekhincp3,Bolognesi1,Bolognesi2,Bolognesi3, flachi2019a, flachi2020} and on circle \cite{monin2015,bolognesi4, fujimori2019, misumi2019}. 
In particular it was shown \cite{Bolognesi1} that in some cases the ground state field configuration must be inhomogeneous namely that expectation values of the fields must depend on the spacial coordinate. 
This observation stimulated search for inhomogeneous solutions of the model on the whole plane without boundaries. Such solutions were firstly constructed in \cite{Nittasol1} using analogy with Gross-Neveu model \citep{Dashen1975a, Feinberg1995a, Feinberg2004} and studied in \cite{Nittasol2, Nittasol3, Gorsky2018, Flachi2019}. In particular it was shown in \cite{Gorsky2018} that the energy of the solution is negative. In other words such inhomogeneous configuration has lower energy than  the homogeneous state. 
However, these solutions break internal $O(N)$ symmetry of the theory so zero modes appear and the solution may suffer from infrared (IR) divergences.  This IR physics probably prevents the system on the whole plane from collapsing into inhomogeneous phase despite of energy consideration. Therefore interpretation of the solutions remains unclear.

The purpose of this note is to expand the analysis of inhomogeneous states  to the case of linear sigma model.
In context of linear models similar solutions were considered in \cite{Abbott, Feinberg1995}. Some of our solution were considered a long time ago in \cite{Feinberg1995}.
We revisit these solutions exploring their properties in greater details. In particular, we carefully compute energy of this solutions 
and find that for some values of parameters such configurations have negative energy similarly to \cite{Gorsky2018}.
We consider various signs of the $n^2$ and $n^4$ terms and study soliton like inhomogeneous field configurations.  At some values of couplings energy changes sign which might indicate phase transition of some kind. 

This note is organized as follows. In Section 2 we review the actions of the considered models and discuss their properties in the large $N$ limit. We introduce the gap equation for the models and determine their spectrum. In Section 3 we discuss inhomogeneous solutions of the gap equations. In Section 4 we calculate the energy of the solutions. In Section 5 the case of finite temperature is considered. The obtained results are summarized in Section 6. 

\section{The models}

The Euclidean action of two dimensional sigma model is 
\begin{equation} \label{eq:action1}
	S = \int d^2 x \, \left(\frac{1}{2} (\partial n_a)^2 + \frac{g}{2N}(n_a^2 - r)^2 \right).
\end{equation}
We use this parameterization of the action in order to emphasize similarity to the nonlinear version of the model. The model consists of $N$ real fields $n_a$, $a=1,\dots,\,N$. Here $r$ is a dimensionless constant, which is traded for a dynamically generated mass scale at quantum level and $g$ is dimensionful coupling constant. In the limit of large positive $g$ potential term in \eqref{eq:action1} leads to the constraint $n_a^2 = r = const$ and we recover the action for the $O(N)$ nonlinear sigma model. Note that the model \eqref{eq:action1} can be considered for the negative values of $g$. It was shown in \cite{Abbott}, that quantum fluctuation stabilize the potential if absolute value $|g|$ of the coupling constant is small enough.
The case $g>0$ corresponds to a classically stable system with spontaneous breaking of the $O(N)$ symmetry. Of cause in two dimensions we expect that the symmetry is restored by quantum fluctuations.

To examine the model in the large $N$ limit
we start by rewriting Euclidean action via the auxiliary field $\lambda$
\begin{equation} \label{eq:action2}
	S = \int d^2 x \, \left(\frac{1}{2} (\partial n_a)^2 + \frac{\lambda}{2}((n_a)^2 - r) - \frac{N \lambda^2}{8g} \right).
\end{equation}
Note that negative sign of the $\lambda^2$ term corresponds to the positive potential.
After integrating out the $n$ fields we obtain the effective action
\begin{equation} \label{eq:effaction}
	S_{eff} = \frac{N}{2} tr \log(-\partial^2 + \lambda)  + 
	\int d^2 x \, \left(\frac{1}{2} (\partial n)^2 + \frac{\lambda}{2}(n^2 - r) - \frac{N \lambda^2}{8g} \right).
\end{equation}
We left one component $n = n_N$ for the consideration of inhomogeneous solutions,
but for a moment we assume that it is equal to zero and that $\lambda$ is constant.

Differentiating the action \eqref{eq:effaction} with respect to $\lambda$ we obtain gap equation 
\begin{equation} \label{eq:gap1}
	\frac{N}{8\pi} \log \frac{M^2}{m^2} - \frac{1}{2}r - \frac{N m^2}{4 g} = 0.
\end{equation}
Here $M^2$ is an UV cutoff, the expectation value of the auxiliary field $\lambda = m^2$ corresponds to the physical mass of the particles and
$\Lambda$ is a dynamically generated scale, introduced via renormalization
\begin{equation} \label{eq:renorm}
	r = \frac{N}{4\pi} \log \frac{M^2}{\Lambda^2}.
\end{equation}
We choose this renormalization procedure to make it similar to the nonlinear sigma model case.
The coupling constant $g$ is not renormalized which is consistent with the diagrammatic expansion in two dimensions.
Now we can explain what we mean by a large coupling: strong coupling limit corresponds to the case $|g|\gg \Lambda^2$.

The final version of homogeneous gap equation reads
 \begin{equation} \label{eq:homgap}
 	m^2 = \frac{g}{2\pi} \log \frac{\Lambda^2}{m^2}.
 \end{equation}
This equation defines the mass of the particles  as a function of the coupling and the mass scale.
In the strong coupling limit $g \gg \Lambda^2$ the solution is  $m = \Lambda$.
In the case of positive coupling constant ($g > 0$) this equation always have a unique solution, 
in the small coupling limit we obtain 

\begin{equation}
	m^2 \approx \frac{g}{2\pi} \log \frac{2\pi\Lambda^2}{g}.
\end{equation}
Thus mass increases with the coupling constant $g$ and can be made arbitrary small.

For the negative coupling the gap equation has solutions only for
\begin{equation}
	\frac{2 \pi \Lambda}{g} < \frac{1}{e}.
\end{equation}
For the allowed values of negative coupling the gap equation \eqref{eq:homgap} has two solutions.
In order to choose one of them we have to calculate the energy density of this states.
To take into account the conformal anomaly we use Pauli-Villars regularization.
Concretely, the vacuum energy density after subtraction of quadratic divergent term is 
\begin{equation}
\epsilon = \frac{Nm^2}{8\pi} \left(1 + \frac{\pi m^2}{g} \right).
\end{equation}
Thus our result for energy is  are slightly different from \cite{Abbott} but the general conclusion that the lowest energy state corresponds to the larger value of mass is the same.
The fact that energy is lower for the larger value of mass can now be confirmed by direct numerical computation for all allowed values of coupling.
In case of the small negative coupling it is obvious from the fact that the energy density is negative thus larger mass leads to large absolute value of energy and lowers energy itself.
In spite of the fact that coupling constant is large when compared to the renormalization scale, it is small compared with the mass.
Therefore stable vacuum corresponds to the weak coupling limit.
In terms of physical mass, coupling constant must satisfy condition 
\begin{equation} \label{restriction}
	\frac{|g|}{2\pi m^2} < 1.
\end{equation}

We also consider the model with quartic interaction and positive mass and coupling constant. Its  Euclidean action is
\begin{equation} \label{positivemod}
S = \int d^2 x\, \left(\frac{1}{2} (\partial n_a)^2  + \frac{1}{2} m_0^2 n_a^2 + \frac{g}{4N}(n_a^2)^2\right).
\end{equation}
Similarly to the previous cases we can rewrite the action via an auxiliary field $\lambda$
\begin{equation}
S = \int d^2 x\, \left( \frac{1}{2} (\partial n_a)^2  + \frac{1}{2} \lambda n_a^2 - \frac{N}{4g} (\lambda - m_0^2)^2 \right).
\end{equation}
Now we can integrate out the scalar fields and obtain the effective action
\begin{equation} \label{eq:effaction2}
S_{eff} = \frac{N}{2} tr \log (-\partial^2 + \lambda) + \int d^2 x\, \left(\frac{1}{2} (\partial n)^2  + \frac{1}{2} \lambda n^2 - \frac{N}{4g} (\lambda - m_0^2)^2\right).
\end{equation}
As usual $n = n_N$.
We consider homogeneous solution for which  $n=0$ and $\lambda$ is constant.
The gap equation is
\begin{equation}
m^2 = m_0^2 + \frac{g}{4\pi} \log \frac{M^2}{m^2}.
\end{equation}
Here $\lambda = m^2$ and $M$ is the UV cutoff.
The solution for the physical mass of the particles $m^2$ is always unique.
Note that if the coupling constant is small we can substitute $m_0$ instead of $m$ in the logarithmic term and thus obtain the usual one loop mass renormalization. However the large $N$ limit allows us to consider coupling of arbitrary strength.

\section{Inhomogeneous solutions}
Now we turn to the inhomogeneous solutions, namely the stationary points of effective action \eqref{eq:effaction} or \eqref{eq:effaction2} for which both fields $\lambda(x)$ and $n(x)$ can depend on spacial coordinate. However, we assume that they do not depend on (Euclidean) time.

Similarly to the \cite{Gorsky2018} we use the ansatz 
\begin{equation} \label{soliton}
	\lambda = m^2 \left(1 - \frac{2}{\cosh^2 mx} \right), n \sim \frac{1}{\cosh mx}.
\end{equation}
Variation of the action \eqref{eq:effaction} with respect to $\lambda$ leads to the gap equation
\begin{equation} \label{inhgap}
	\frac{N}{4\pi} \sum_n \frac{|f_n(x)|^2}{2 E_n} + \frac{1}{2} (n^2 -r) - \frac{N}{4g} \lambda = 0.
\end{equation}
	Here summation is over eigenfunction of differential operator $-\partial_x^2 + \lambda(x)$:
	\begin{equation}
	(-\partial_x^2 + \lambda(x))f_n(x) = E_n^2 f_n(x).
	\end{equation}
	Eigenfunctions for the field configuration \eqref{soliton} are
	\begin{equation} \label{modes}
	f_k(x) = \frac{-\im k + m}{\sqrt{k^2 + m^2}} e^{\im k x}, \; E_n^2 = k^2 + m^2.
	\end{equation}
	Also there is a zero mode $f_0 \sim 1/(\cosh mx)$. It corresponds to the rotations of the solution in the internal space, therefore we explicitly exclude this mode from summation in \eqref{inhgap}. In partition function integration over the zero modes yields the volume of the moduli space of the solution that is volume of $N-1$ dimensional sphere.  
	
	Varying the action with respect to $n$ we obtain equation
	\begin{equation}
	(-\partial_x^2 + \lambda(x)) n(x) = 0,
	\end{equation}
	which is satisfied by \eqref{soliton} automatically.
	
After substitution \eqref{modes} in the gap equation \eqref{inhgap} we find that coordinate independent terms cancel due to \eqref{eq:homgap}
and inhomogeneous part gives the amplitude of the $n$ field:
\begin{equation}
	n^2 = \frac{N \lambda}{2g} - N \int \frac{dk}{4\pi} \left( \frac{1}{\sqrt{k^2 + m^2}} - \frac{1}{\sqrt{k^2 + \Lambda^2}}\right)
	+ \frac{N}{4\pi} \int_{-\infty}^{\infty} \frac{dk}{\sqrt{k^2 + m^2}} \frac{m^2}{k^2 + m^2} \frac{1}{\cosh^2 mx}.
\end{equation}
After straightforward integration we obtain
\begin{equation} \label{condensate}
	n^2  = \frac{N}{2\pi} \left(1 - \frac{2\pi m^2}{g}\right)\frac{1}{\cosh^2 mx}.
\end{equation}

Note that for negative coupling we always have a solution and the only condition is \eqref{restriction}.
For the positive coupling we have a nontrivial restriction $n^2 \ge 0$ or 
\begin{equation} \label{cond}
	\frac{g}{2 \pi m^2} > 1.
\end{equation}
The solution exists only for strong enough coupling. 

Now we turn to the model with action \eqref{eq:effaction2}.
The ansatz \eqref{soliton} is the same as previously.
The gap equation reads as 
\begin{equation} \label{inhgap2}
	\frac{N}{4\pi} \sum_n \frac{|f_n(x)|^2}{2 E_n} + \frac{1}{2} n^2 - \frac{N}{2g}( \lambda -m_0^2) = 0.
\end{equation}
Therefore we can calculate $n^2$. The result is given by the same equation \eqref{condensate}. 
Again the solution exists only when condition \eqref{cond} is satisfied. 
However one should remember that in this case the meaning of the coupling constant $g$ is different.

\section{Energy of the solutions}

Now we are going to calculate the energy of the solution found in previous section. We introduce large but finite time cutoff $\beta$ and calculate the regularized Euclidean effective action $S_{reg}$ of this solutions. 
To deal with divergences we use Pauli-Villars regularization for calculating the functional determinants and subtract the action for the homogeneous solution. The energy is $E = S_{reg}/\beta$.

The Pauli-Villars regularized effective action for the model \eqref{eq:action2} is 
\begin{equation} \label{eq:regeffaction}
	S_{reg} = \frac{N}{2} \sum_i C_i \tr \log(-\partial^2 + \lambda +M_i^2)  + 
	\int d^2 x \, \left(\frac{1}{2} (\partial n)^2 + \frac{\lambda}{2}(n^2 - r) - \frac{N \lambda^2}{8g} \right).
\end{equation}
The summation is over regulator fields, $M_i$, $i=0,\,1,\,2$, are regulator masses and $C_i$ are coefficients satisfying 
$$
\sum_i C_i = 0; \; \sum_i C_i M_i^2 =0; \; C_0 = 1, \; M_0 = 0.
$$
Coupling constant $r$ can be expressed in terms of regulators' masses as
\begin{equation}
	r = - \frac{N}{4\pi} \sum_{i=1,\,2} \log \frac{M_i^2}{\Lambda^2}.
\end{equation}
Subtracting from \eqref{eq:regeffaction} similar expression for homogeneous configuration  we obtain energy
\begin{multline} \label{calc_energy}
	E = \frac{N}{2}\int \frac{d \omega}{2 \pi}  \sum_i C_i \log(\omega^2 + M_i^2) + \frac{N}{2} \int \frac{d \omega}{2 \pi} \sum_n \sum_i C_i \log\frac{\omega^2 + E_n^2 + M_i^2}{\omega^2 + E_{n0}^2 + M_i^2} + \\
	 \int dx\, \left(\frac{1}{2} (\partial_x n)^2 + \frac{1}{2}\lambda n^2 - \frac{r}{2}(\lambda - m^2) - \frac{N}{8g}(\lambda^2 - m^4)\right).
\end{multline}
The first term is the contribution from the zero mode, the second term comes from continuous spectrum and the last term is from classical part of the action. To calculate the second term one should remember that eigenvalues in continuous spectra of homogeneous and inhomogeneous configurations are the same but eigenvalue densities are different. The difference as function of momentum $k$ is 
\begin{equation*}
	\rho(k) = \frac{1}{\pi} \frac{\mathrm{d} \delta}{\mathrm{d} k} = - \frac{2m}{\pi (k^2 + m^2)}.
\end{equation*}
Otherwise calculation is straightforward.
The final expression for energy is
\begin{equation}
	E = -\frac{Nm}{\pi} \left(1 + \log \frac{\Lambda}{m} - \frac{\pi m^2}{3 g} \right)=
	 -\frac{Nm}{\pi} \left(1  +\frac{2\pi m^2}{3 g} \right).
\end{equation}
Here the first term is the conformal anomaly contribution, the second is due to the renormalization 
of the coupling constant $r$ and the third is a contribution of $\lambda^2$ term.
In the last transformation the homogeneous gap equation was used.
This expression is  negative when $g>0$.
However in case $g<0$ the energy can change sign. The homogeneous ground state is stable when $2 \pi m^2/ |g| > 1$
and the  coupling constant value $|g| = 2 \pi m^2/3 $
at which energy changes sign is allowed. 
Therefore at small negative coupling solitons behave as usual excitations  whith positive energy. 
 
For the model \eqref{positivemod} calculation can be performed in similar way. The result is
\begin{equation}
E = - \frac{N m}{\pi} - \frac{4 m^3}{3g}N.
\end{equation}
Clearly the energy is always negative.

\section{Finite temperature}

In this section we discuss the model with the action \eqref{eq:action2} at finite temperature. For simplicity we restrict analysis only to the case of large $g$, namely we consider only the case of the nonlinear sigma model.
The effective action can be calculated in the same way as in zero temperature case. The only difference is that the trace should be taken over periodic fields with period $\beta = 1/T$ in Euclidean time. Therefore instead of integration over all frequencies we calculate the sum over Matsubara frequencies $ \Omega_n = 2 \pi T n $.

Firstly we consider homogeneous saddle points of the action.
The gap equation yields
\begin{equation}
	NT \int \frac{dk}{2\pi} \sum_{n \in \mathbb{Z}}
		\frac{1}{k^2 + m^2 + \Omega_n^2} -r = 0,
\end{equation}
or after summation over frequencies
\begin{equation}
	\frac{N}{4\pi} \int_0^{\infty} \df k \left( 
	\frac{\coth\left( \frac{1}{2T} \sqrt{k^2 +m^2}\right)}{\sqrt{k^2 + m^2}} -
	\frac{1}{\sqrt{k^2+ \Lambda^2}} \right) = 0.
\end{equation}
The second term here is the integral representation of coupling constant $ r $. From this equation we can determine mass as a function of temperature.
In the low temperature limit $T \ll \Lambda$ we have $m \approx \Lambda$ and
for $ T \gg \Lambda $ 
the solution of  is 
\begin{equation}
	m = \frac{\pi T} {\log (\kappa T/\Lambda) }, \; \kappa \approx 7.08.
\end{equation} 

Now consider soliton solution \eqref{soliton}.
From the gap equation we obtain
\begin{equation}
n^2 = \frac{N}{4\pi} \int \frac{m^2 \df k}{\left(k^2 + m^2\right)^{3/2}}  \coth\left( \frac{\sqrt{k^2 + m^2}}{2T}\right) \frac{1}{\cosh^2 mx} 
= \frac{NA}{\cosh^2 mx}.
\end{equation}
The last equation is a definition of the amplitude $A$ of the condensate. At high temperature we have
\begin{equation}
A \approx \frac{T}{4m} \left(1+ \frac{m^2}{6 T^2} \right).
\end{equation}

Now we turn to the calculation of the free energy of the soliton. The free energy is connected with the regularized action\eqref{eq:regeffaction} as $E = T S_{reg}$. The only subtlety of the calculation is that we have to take care of the zero mode in the determinant term of effective action.
If we treat the zero mode in Gaussian approximation as all other modes, it will lead to an infrared divergence. The reason is that the $n$ field in the solution \eqref{soliton} breaks internal $O(N)$ symmetry in the model which results in orientational zero modes. As 
typical for calculations with solitons we have to integrate over moduli space rather than use Gaussian approximation.

The moduli space of the soliton is the $N-1$ dimensional sphere $S^{N-1}$ (translational mode does not contribute to the effective action in the leading order of $1/N$ expansion). Thus moduli space dinamics can be represented by a time-dependent unit vector $l^a(t)$ and the soliton configuration is 
$$
	n^a(x,\,t) = n(x) l^a(t).
$$
Here $n(x)$ is previously found solitonic solution. Corresponding effective action for $l^a$ is 
\begin{equation}
	S_1 = \frac{1}{2} \int \df x\, n^2(x) \int \df t \,\dot{l}^a(t)^2 = \frac{M}{2} \int \df t\, \dot{l}^a(t)^2; \quad M = \frac{2 N A}{m}.
\end{equation}
This action is formally the same as for a non-relativistic particle of the mass $M$ on a $N-1$ dimensional sphere with unit radius. For this system the separation of energy levels is of order $1/M \sim m/N$. Therefore partition function can be calculated classically if the temperature is not too small ($T \gg m/N$). This assumption is reasonable in the large $N$ limit. Classical partition function is
\begin{equation}
	Z_1 =\frac{1}{(2 \pi)^{(N-1)/2}} S^{N-1} \left(2\pi M T\right)^{(N-1)/2} \approx \left( \frac{2 e A T}{m}\right)^{N/2}.
\end{equation}
Here $S^{N-1}$ is the area of the sphere. Thus zero mode contribution to the free energy is $-T \log Z_1$.

After straightforward computation we find the free energy of the soliton
\begin{multline}
	E = -\frac{Nm}{\pi} - \frac{Nm}{\pi} \log \frac{\Lambda}{m}
		- \frac{2NT}{\pi} \int_0^{\infty} \frac {mdk}{m^2 + k^2} 
		\log\left( 1 - \exp \left(- \frac{\sqrt {k^2 +m^2}}{T}\right)\right) -\\- \frac{NT}{2} \log\frac{2 e A T}{m} \approx mN \left(\frac{1 - \log 2}{2\pi} \log \frac{T}{\Lambda} + C -\frac{1}{4 \log(\kappa T/\Lambda)} \right);\; C \approx 0.0945.
\end{multline}
Here the first term is due to the zero temperature fluctuations, the second term comes from the coupling constant renormalization.
The third term is the contribution of thermal excitations.
The last term comes from the zero mode.

The free energy is negative at small temperatures and increases with $T$. On the other hand, in the high temperature limit free energy becomes positive due to the thermal excitaitions. At the point $T \approx 1.044 \Lambda$, which can be found numerically, energy changes sign. This observation suggests that the model might undergo a phase transition of some kind. 

\section{Conclusions}

To sum up, we considered inhomogeneous solutions in linear sigma model and investigated different signs of the coupling constant.  
We schematically will write the signs of the different terms in this cases and summarize the corresponding solutions.
Properties of the soliton for different signs of mass squared and coupling constant are summarized in the table \ref{table}. 

\begin{table}[t]
\centering
\begin{tabular}{|c|c|}
\hline
Potential & Properties of the soliton \\ \hline
$m^2>0,\,g>0$ & Soliton exists at large coupling but disappears at weak coupling.\\ \hline
$m^2<0,\, g>0$ & Soliton exists only at strong coupling, energy is always negative. \\ \hline
$m^2>0,\, g<0$ & Soliton always exists but its energy changes sign. \\ 
\hline
\end{tabular}
\caption{Properties of the soliton in model with potential $V= m^2 n^2 + g n^4$. }
\label{table}
\end{table}

One can note that only the sign of the coupling constant is important, behavior of the models with different signs of $m^2$ is very similar. Disappearance of the solitons at positive coupling and changes in the sign of the energy might indicate phase transition of some kind. Physical meaning of such a transition, if exists, is very unclear and calls for an explanation. We postpone this task for a future work.

This solutions are similar to discussed in \cite{Feinberg1995}. However, there are some important differences. 
Firstly, in \cite{Feinberg1995} only the dynamics of $\lambda$ field was considered, symmetry breaking scalar condensate $n$ was not introduced. Therefore our saddle point equations are rather different from \cite{Feinberg1995}.
Secondly, in \cite{Feinberg1995} anomalious contribution to the energy density was not taken into account. Therefore energy is found to be positive and solitons were interpreted as some excited (metastable) states of the theory. 

The nonlinear sigma model at finite temperature was also discussed. We found that soliton always exists, but its energy  becomes positive when the temperature is high enough. Therefore another phase transition is possible.

\section*{Acknowledgment}
The author is grateful to Alexander Gorksy for suggesting this problem and numerous discussions. The work of A.P. was supported by Basis Foundation fellowship and RFBR grant 19-02-00214.

\end{document}